\documentclass[final,3p,times]{elsarticle}

\usepackage{amssymb}
\usepackage{amsmath}
\usepackage{lineno}  
\usepackage{xcolor}
\usepackage{hyperref}  
\usepackage{overpic}   
\usepackage{wrapfig}   
\usepackage{upgreek}
\usepackage{multicol}

\newcommand{\pT}{$p_{\rm T}$}
\newcommand{\Bzero}{$\rm B^{0}$}
\newcommand{\Dzero}{$\rm D^{0}$}
\newcommand{\jpsi}{$\mathrm{J}/\psi$}
\newcommand{\GevPerc}{GeV/$c$}

\journal{Journal of Subatomic Particles and Cosmology}
\biboptions{sort&compress}

\begin{document}

\begin{frontmatter}

\title{Heavy-flavour production and correlations in pp collisions:
precision tests of pQCD and hadronisation with ALICE}

\author[iitb]{Deependra Sharma\corref{cor1}}
\ead{deep.phy@cern.ch}
\cortext[cor1]{For the ALICE Collaboration}

\affiliation[iitb]{organization={Indian Institute of Technology Bombay},
             city={Mumbai},
             postcode={400076},
             country={India}}

\begin{abstract}
  Heavy quarks (charm and beauty) are predominantly produced in hard partonic scatterings,
  making their cross sections in proton--proton (pp) collisions calculable in
  perturbative quantum chromodynamics (pQCD) and thus providing stringent
  tests of pQCD. Furthermore, the associated production of two charm 
  hadrons in a single collision offers a sensitive probe of multiparton interaction dynamics, 
  distinguishing between single parton scattering (SPS) and double parton scattering (DPS) processes.

  In this contribution, preliminary measurements of prompt D-meson production are reported, together with the final 
  results of the \Bzero-meson production cross section down to \pT \space = 1 \GevPerc \ at midrapidity. 
  The rapidity dependence of B-meson production is investigated by computing the 
  ratio with respect to LHCb measurements at forward rapidity. The associated 
  production of \Dzero--\jpsi \space pairs in pp collisions at $\sqrt{s} = 13.6$\ TeV is
  presented as well, where \Dzero \ mesons are reconstructed at midrapidity, while 
  \jpsi \space candidates are measured at forward rapidity. 
  These measurements are compared with pQCD calculations and 
  phenomenological models, providing crucial constraints on heavy-quark production, 
  hadronisation, and multiparton interaction dynamics.
\end{abstract}


\end{frontmatter}
\vspace*{-\intextsep}%
\section{Introduction}
\label{sec:intro}

Heavy quarks, charm ($m_{\rm c} \approx 1.3$\ GeV/$c^{2}$) and beauty
($m_{\rm b} \approx 4.2$\ GeV/$c^{2}$), are predominantly produced in hard partonic
scatterings and their large masses ensure $\alpha_{\rm s}(m_\mathrm{Q})\ll 1$,
making their production cross sections calculable in pQCD via collinear
factorisation. The measurements of heavy-flavour hadron production provide precise tests of pQCD-based models, 
constrain parton distribution functions (PDFs) and fragmentation functions (FFs), and serve as essential baselines
for heavy-ion studies. The ALICE Run~3 dataset
enables measurements of unprecedented statistical precision, which allow probing not only kinematic regions 
where theoretical uncertainties are large, but also those where previous experimental measurements were limited.




\section{Prompt D$^{0}$-meson production}
\label{sec:D0}
\begin{wrapfigure}[23]{l}{0.48\columnwidth}
  \centering
  \includegraphics[width=\linewidth,keepaspectratio]{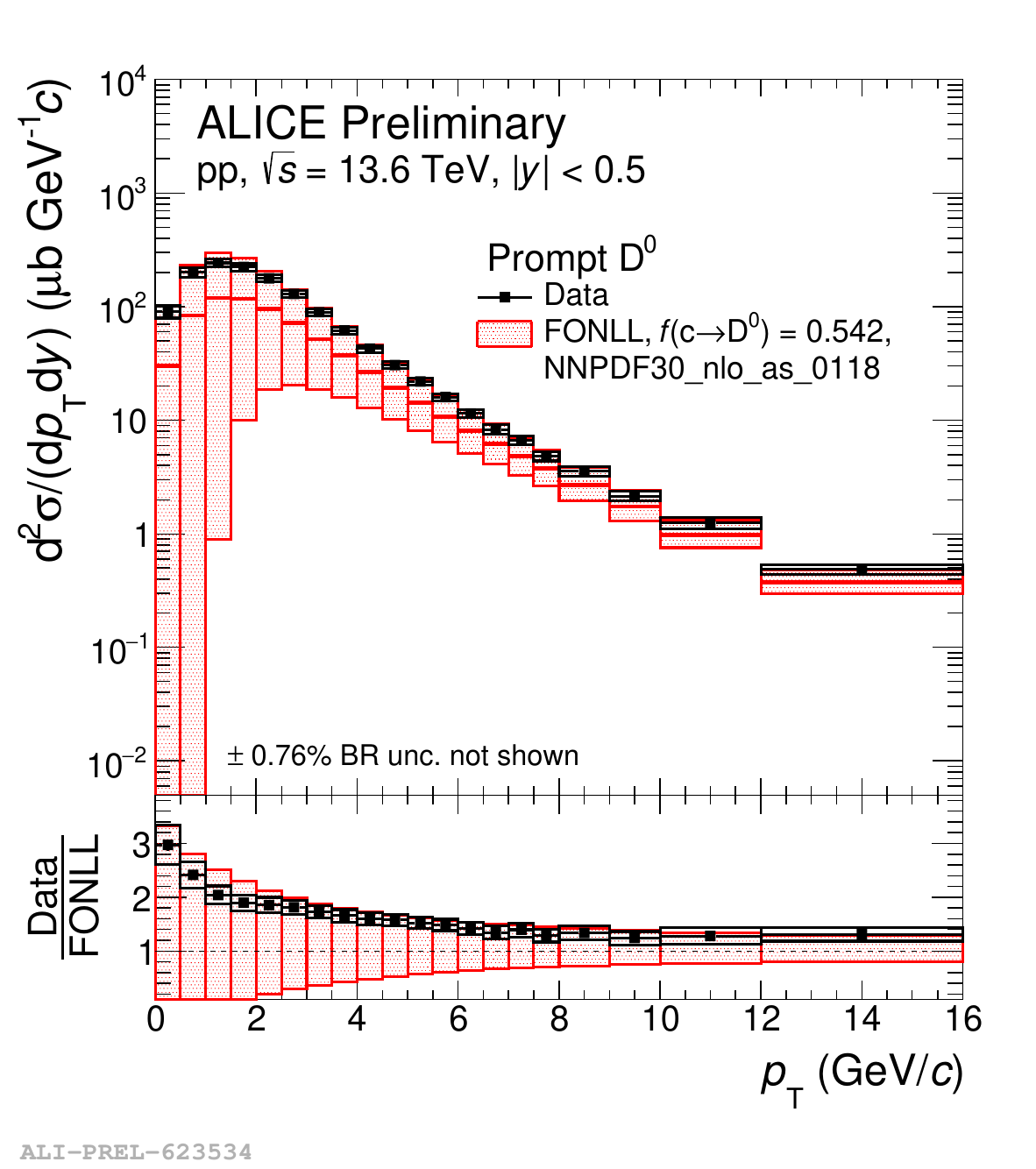}
  \caption{The prompt \Dzero -meson production cross section measured in pp collisions at $\sqrt{s} = 13.6$\ TeV
  compared with theoretical predictions from FONLL ~\cite{Cacciari:1998it}. 
   The lower panel shows the data-to-theory ratio.}
  \label{fig:D0}
\end{wrapfigure}

Prompt \Dzero -meson production is measured by reconstructing the \Dzero \ mesons through the hadronic decay 
channel D$^{0} \to {\rm K}^{-}\uppi^{+}$ (and its charge conjugate) in the kinematic interval of $0 < p_{\rm T} < 16$\ GeV/$c$
and $|y| < 0.5$. The measurement is performed using a sample of pp collisions at $\sqrt{s} = 13.6$\ TeV, 
corresponding to an integrated luminosity of $\mathcal{L}_{\rm int} = 5$\,pb$^{-1}$.
The \Dzero \ mesons are selected by applying a boosted decision tree (BDT) algorithm trained to separate prompt \Dzero \ mesons from 
non-prompt \Dzero \ and background candidates. The signal is extracted via an invariant-mass fit of the selected K$\uppi$ pairs in various 
\pT\ intervals.
The remaining non-prompt component, not rejected by the BDT selections, is removed by using a data-driven method as discussed in Ref.~\cite{DataDriven}.
Figure~\ref{fig:D0} reports the prompt \Dzero-meson production cross section compared with theoretical calculations from FONLL~\cite{Cacciari:1998it}.
The data are in agreement with model predictions and lie on the
upper edge of the model uncertainty band. The theoretical uncertainties are
much larger than the experimental ones at low \pT, so the measurement provides constraints on pQCD calculations.

\section{B$^{0}$-meson production}
\label{sec:B0}
\begin{figure}[!b]
\centering
\begin{minipage}[t]{0.48\columnwidth}
  \centering
  \includegraphics[width=\linewidth,keepaspectratio, height=1.09\columnwidth]{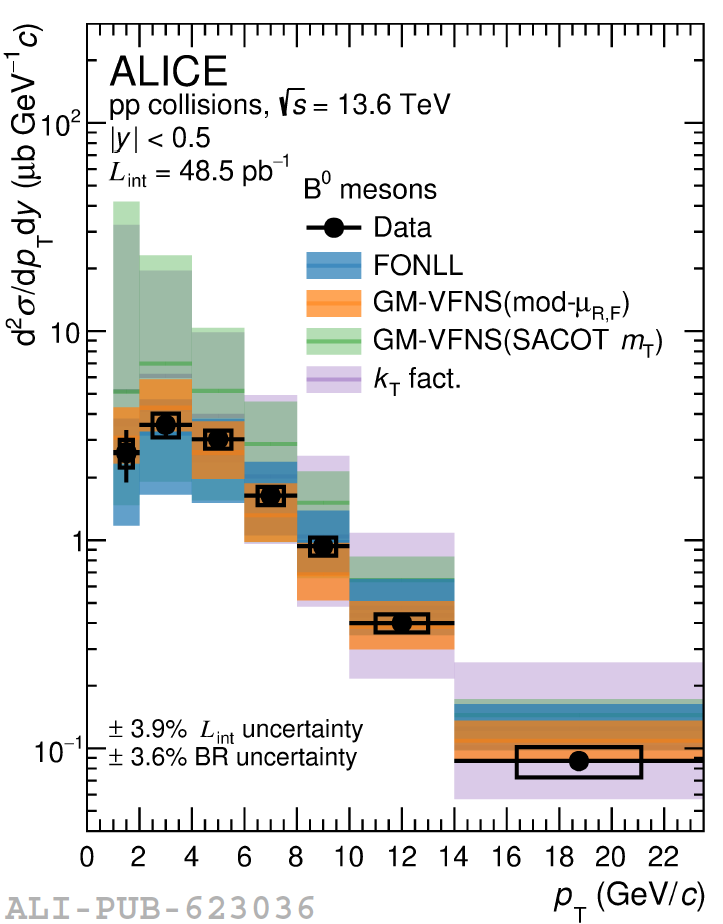}
  \caption{The \Bzero -meson production cross section at midrapidity measured in pp collisions at $\sqrt{s} = 13.6$ \,TeV compared with
  pQCD-based models 
  ~\cite{Cacciari:1998it, Benzke:2019VFNS, Helenius:2023SACOT, Barattini:kT}.
    }
  \label{fig:B0xsec_pQCD}
\end{minipage}
\hfill
\begin{minipage}[t]{0.48\columnwidth}
  \centering
  \includegraphics[width=\linewidth,keepaspectratio, height=1.09\columnwidth]{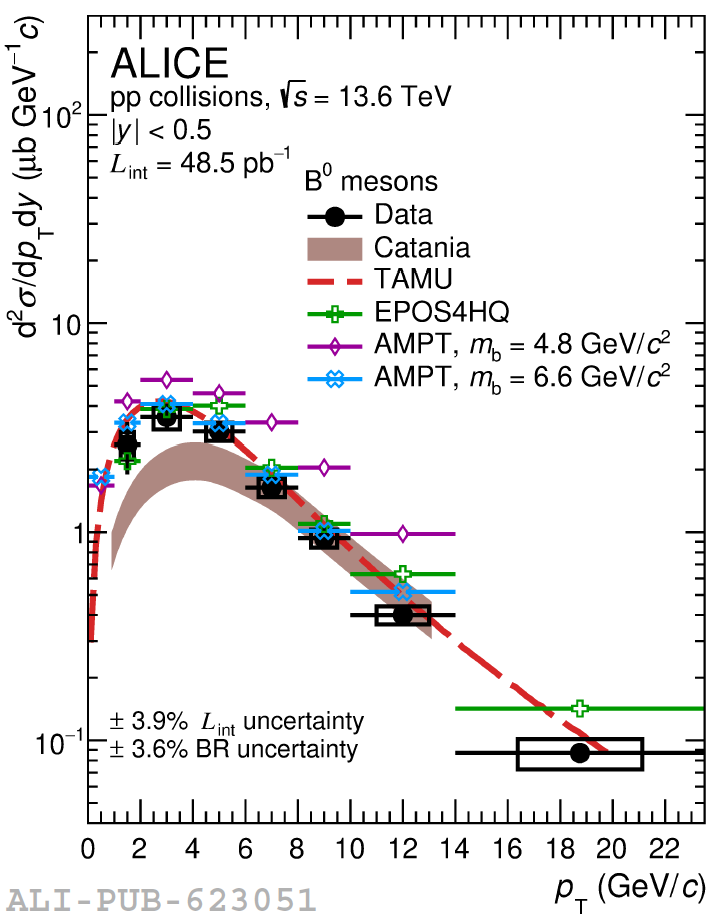}
  \caption{The \Bzero -meson production cross section at midrapidity measured in pp collisions 
  at $\sqrt{s} = 13.6$\ TeV compared with transport-model
  calculations~\cite{Minissale:2021Catania,He:2023TAMU,Zhao:2024EPOS4HQ,He:2026AMPT}.}
  \label{fig:B0xsec_pheno}
\end{minipage}
\end{figure}



The \Bzero -meson production cross section is measured for the first time down to $p_{\rm T} = 1$\ \GevPerc \
at midrapidity in pp collisions at $\sqrt{s} = 13.6$\,TeV by the ALICE Collaboration~\cite{ALICE:2025B0}. The measurement 
is based on pp collision data collected in 2023 and 2024, corresponding to an integrated luminosity of 
$\mathcal{L}_{\rm int} = 48.5$\,pb$^{-1}$.
The measurement exploits the hadronic decay channel ${\rm B}^{0}\to{\rm D}^{-}\uppi^{+}\to\uppi^{-}{\rm K}^{+}\uppi^{-}\uppi^{+}$ to
reconstruct the \Bzero -meson invariant mass distribution.

An offline trigger selection strategy is developed to select events containing decays of beauty hadrons into a 
D meson and a pion. The \Bzero \ candidates are further selected by using a BDT model trained on candidates from triggered 
events to separate signal \Bzero \ candidates from the combinatorial background. The \Bzero \ raw yields are extracted 
via unbinned-likelihood fits to the invariant mass 
distributions of the selected \Bzero \ candidates. Background contributions from partially reconstructed or misreconstructed 
beauty-hadron decays
are considered in the fit by including the template distributions obtained from Monte Carlo simulations.


The \Bzero -meson production cross section  is reported in Figs.~\ref{fig:B0xsec_pQCD} \ and ~\ref{fig:B0xsec_pheno}.
In Fig.~\ref{fig:B0xsec_pQCD}, the measurement is compared with pQCD-based calculations. All the models are compatible 
with the measurement within uncertainties. FONLL~\cite{Cacciari:1998it} describes the data well over the full \pT\ interval.
GM-VFNS (mod-$\upmu_{\mathrm{R,F}}$)~\cite{Benzke:2019VFNS} tends to underestimate the measurement in the interval 4 < \pT\ < 10 GeV/$c$,
while GM-VFNS (SACOT-$m_{\rm T}$)~\cite{Helenius:2023SACOT} tends to overestimate the data. 
This comparison between the two pQCD calculations (mod-$\upmu_{\mathrm{R,F}}$, SACOT-$m_{\rm T}$) tests which one better handles 
the divergences at 
low $p_{\rm T}$ and the resulting uncertainties.
The $k_{\rm T}$-factorisation-based model~\cite{Barattini:kT} is in agreement with the data for $p_{\rm T} > 4$\ \GevPerc, though with larger uncertainties than the other calculations.

In Fig.~\ref{fig:B0xsec_pheno}, the \Bzero -meson production cross section is compared with transport models.
The TAMU model~\cite{He:2023TAMU}, which considers statistical hadronisation of beauty quarks, describes the data within uncertainties.
The Catania model~\cite{Minissale:2021Catania}, assuming the formation of a quark--gluon plasma in pp collisions and including \pT-dependent 
coalescence plus fragmentation, underestimates the data at low \pT. Both models use the \pT -differential b-quark 
distribution from FONLL. The EPOS4HQ~\cite{Zhao:2024EPOS4HQ}, which employs the same hadronisation scheme as Catania, agrees with the data only 
up to approximately $10$~\GevPerc\ with the b$\mathrm{\bar{b}}$-quark distribution obtained from next-to-leading-order (NLO) calculations.
The AMPT~\cite{He:2026AMPT}, a multiphase transport model, describes the data with a beauty-quark mass of $m_{\rm b}=6.6$\ GeV/$c^2$,
but overestimates the measurement when the mass is set to $m_{\rm b}=4.8$\ GeV/$c^2$. 

\begin{figure}[!b]
  \centering
  \includegraphics[width=0.9\columnwidth,keepaspectratio]{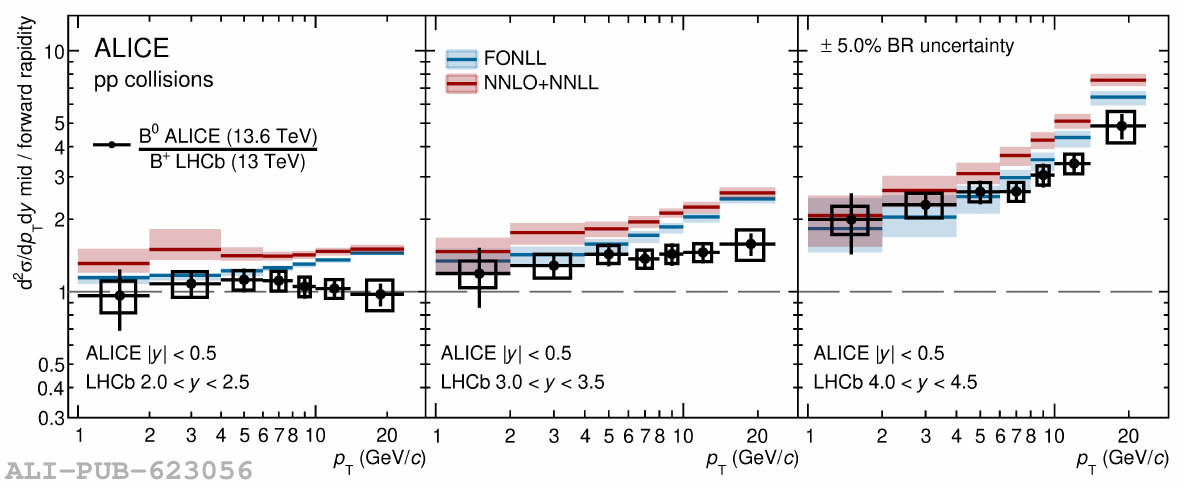}
  \caption{The production ratios of B$^{0}$ (ALICE, mid-$y$) to B$^{+}$ (LHCb, forward $y$) vs.\ \pT\ compared with FONLL 
  and NNLO+NNLL calculations based models
  ~\cite{LHCb:Bplus,Czakon:2025NNLO}.
  }
  \label{fig:B0ratio}
\end{figure}

\indent The rapidity dependence of \Bzero-meson production
is presented in Fig.~\ref{fig:B0ratio}, where the ratio is evaluated relative to the isospin doublet $\rm B^{+}$ production measured by the 
LHCb experiment at forward rapidity at $\sqrt{s} = 13$ \,TeV~\cite{LHCb:Bplus}.
When considering the 2 <\ $y$ \ < 2.5 rapidity interval in the denominator, the ratios do not show a significant 
\pT\ dependence, while an increase with increasing \pT\ is found for larger rapidities in the denominator. 
The measurement is in agreement within uncertainties with FONLL and 
next-to-next-to-leading order + next-to-next-to-leading logarithm (NNLO+NNLL) calculations~\cite{Czakon:2025NNLO}. 


\section{Associated production of D$^{0}$ and J/$\psi$}
\label{sec:DPS}

Associated production refers to the simultaneous production of two particles in a single pp collision.
Such final states can be produced either through a single hard scattering between two partons, known 
as single parton scattering (SPS), or through two independent hard scatterings occurring within the same 
collision, known as double parton scattering (DPS). More generally, DPS represents the simplest 
realisation of Multiple Parton Interactions (MPI). At LHC energies, the high gluon density at low 
Bjorken-$x$ increases the probability of MPI, which can involve hard, semi-hard, and soft scatterings. 
As a result, the DPS contribution to the production cross section can become significant compared to 
the SPS contribution.
Under the assumption that the two hard scatterings are independent, the DPS cross section factorises as
$\sigma_{\rm DPS}({\rm X+Y}) = \sigma({\rm X})\,\sigma({\rm Y})/\sigma_{\rm eff}$ \cite{DPS:MarkusDiehl},
where $\sigma_{\rm eff}$ encodes the transverse parton overlap and is
assumed to be process-independent, and $\sigma({\rm X})$ and $\sigma({\rm Y})$ are the production cross sections of 
the individual particles. Measuring $\sigma_{\rm eff}$ in different
final states provides a test of its universality.

\begin{wrapfigure}[22]{r}{0.5\columnwidth}
\centering
  \includegraphics[width=0.5\columnwidth,keepaspectratio]{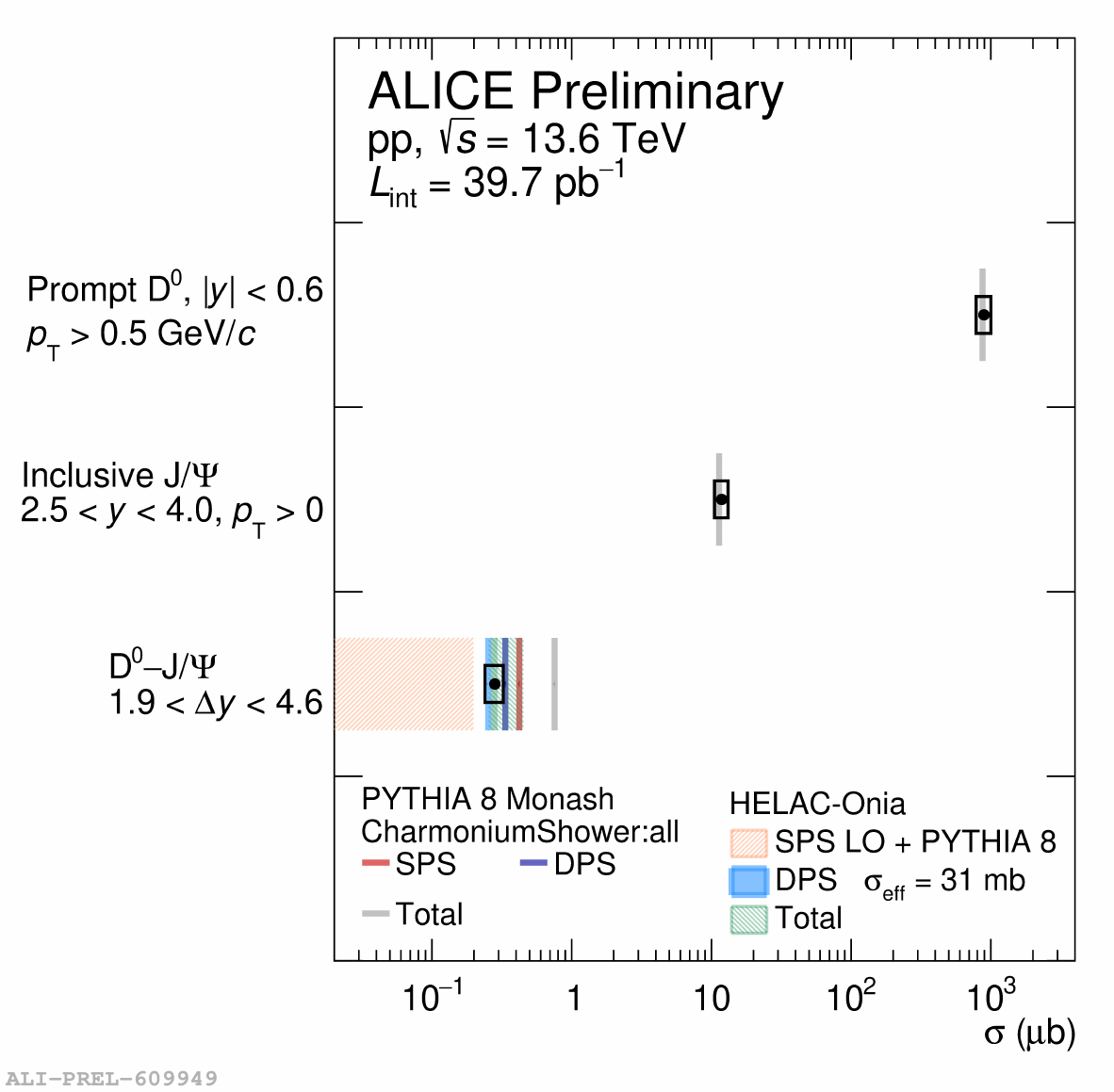}
\caption{The production cross sections of prompt \Dzero, inclusive \jpsi,
and associated D$^{0}$--J/$\psi$ compared with PYTHIA~8 and HELAC-Onia
predictions~\cite{Skands:2014PYTHIA8,Shao:2012HELAC}.
The result is consistent with $\sigma_{\rm eff} = 31$\ mb.} \label{fig:D0Jpsi}
\end{wrapfigure}

The associated \Dzero--\jpsi \ production cross section is measured in
pp collisions at $\sqrt{s}=13.6$\ TeV with
$\mathcal{L}_{\rm int} = 39.7$\,pb$^{-1}$. The \jpsi \ is reconstructed
at forward rapidity ($2.5 < y_{\upmu\upmu} < 4.0$) via $\upmu^+\upmu^-$;
prompt \Dzero \ mesons are reconstructed at midrapidity
($|y_{\uppi{\rm K}}| < 0.6$, $p_{\rm T} > 0.5$\ \GevPerc). The raw
yield is extracted from a two-dimensional fit to the
invariant-mass distributions in the
$M(\mathrm{K}^{-}\uppi^{+})$ vs. $M(\upmu^{+}\upmu^{-})$ plane. Results are compared with theoretical predictions in 
Fig.~\ref{fig:D0Jpsi}. PYTHIA~8~\cite{Skands:2014PYTHIA8} (Monash + CharmoniumShower) overestimates the associated 
production cross section
despite reproducing the single-particle cross sections.
HELAC-Onia~\cite{Shao:2012HELAC} (NLO SPS + PYTHIA~8 DPS) reproduces the
measurement with $\sigma_{\rm eff} = 31$\,mb. 
The large uncertainties in the SPS contribution
may be reduced by higher-order calculations. Both models predict a significant DPS contribution of the 
same order as the SPS contribution, indicating that the associated production of \Dzero--\jpsi \ pairs is a
promising channel to study MPI dynamics.



\section{Summary}
\label{sec:summary}

Recent measurements of heavy-flavour production in pp collisions at
$\sqrt{s} = 13.6$\,TeV performed by the ALICE Collaboration have been reported:
(i) The prompt \Dzero -meson production cross section is measured at midrapidity in the \pT \ interval 
$0<p_{\rm T}<16$~\GevPerc \ with small statistical and systematic uncertainties and is described by FONLL calculations.
(ii) The \Bzero-meson production cross section is measured for the first time down to $p_{\rm T}=1$\ \GevPerc \ at 
midrapidity and is described by various models.
(iii) The associated D$^0$--J/$\psi$ production measurement is consistent with
$\sigma_{\rm eff} = 31$\,mb, constraining DPS/MPI dynamics.

\begingroup
\fontsize{7.8pt}{0.5pt}\selectfont
\setlength{\bibsep}{0.5pt}
\bibliographystyle{elsarticle-num}
\bibliography{sqm2026V5Title}
\endgroup

\end{document}